# Modified black hole solution with a background Kalb–Ramond field

L. A. Lessa[1], J. E. G Silva[2,a], R. V. Maluf[1], C. A. S. Almeida[1]

[1] Departamento de Física, Universidade Federal do Ceará (UFC), Campus do Pici, C.P. 6030, Fortaleza, CE 60455-760, Brazil
[2] Universidade Federal do Cariri (UFCA), Av. Tenente Raimundo Rocha, Cidade Universitária, Juazeiro do Norte, Ceará CEP 63048-080, Brasil

**Abstract** We study the gravitation effects on a static and spherically symmetric spacetime due to the vacuum expectation value (VEV) of a Kalb–Ramond field. The Kalb–Ramond VEV is a background tensor field which produces a local Lorentz symmetry breaking (LSB) of spacetime. Considering a non-minimal coupling between the Kalb–Ramond (VEV) and the Ricci tensor, we obtain an exact parameter-dependent power-law modified black hole. For a particular choice of the LSB parameter, the Lorentz violation produces a solution similar to the Reissner–Nordstrom, despite the absence of charge. The near-horizon geometry is modified by including a new inner horizon and shifting the Schwarzschild horizon. Asymptotically, the usual Minkowski spacetime with a background tensor field is recovered. The vacuum configurations are studied considering the energy conditions and the Lorentz violating source properties. By means of the mercury perihelion test, an upper bound to the local Lorentz violation (LV) is obtained, and its corresponding effects on the black hole temperature is investigated.

## 1 Introduction

The search for reminiscent quantum gravity effects at low energy regime has attracted attention over the last decades. Some models in string theory [1–3], Very special relativity [4,5], Doubly special relativity [6], noncommutative spacetime [7], Horava gravity [8] and Loop quantum gravity [9,10] among other, assume that the Lorentz symmetry might be broken in the gravitational UV regime.

A mechanism for the local Lorentz violating is provided by a spontaneous symmetry breaking potential due to self-interacting tensor fields [1–3,11]. The vacuum expectation value (VEV) of these tensor fields yields to background tensor fields, which by coupling to the Standard Model (SM) fields violate the particle local Lorentz symmetry [11].

The simplest self-interacting tensor field is a vector field, the so-called bumblebee [11,12] whose VEV defines a privileged direction in spacetime. In flat spacetime, the bumblebee fluctuations over the VEV have two massless modes, known as the Nambu–Goldstone modes and one massive or Higgs mode [12,13]. In cosmology, the effects of the bumblebee field on the universe expansion were analysed [14]. In addition, static and spherically solutions were obtained considering a localized source and non-minimal couplings between the Ricci tensor and the VEV vector. Imposing a constancy on the squared norm of the VEV vector, the authors found a modified black hole solution keeping invariant the event horizon [15]. By assuming a covariant constant VEV, the authors found a black hole solution with an interesting modified event horizon [16]. A vacuum Kerr-like solution was also found in Bumblebee gravity [19,20]. The bumblebee vacuum vector also allows exotic solutions, such as the wormhole solutions [17,18]. The effects of the modified geometry due to the bumblebee VEV on the black hole thermodynamics properties were considered in Ref. [35].

In this work, we consider that the Lorentz symmetry breaking (LSB) is driven by a self-interacting antisymmetric 2-tensor, $B_{\mu\nu}$, the so-called Kalb–Ramond (KR) field [21]. Likewise the graviton and dilaton, the Kalb–Ramond field, arises in the spectrum of the bosonic string theory [21]. Assuming that the potential $V$ has a nonzero vacuum expectation value $b_{\mu\nu}$, such antisymmetric background tensor can be decomposed in two spacelike vectors and one timelike vector, resembling the electromagnetic tensor $F_{\mu\nu}$ decomposition [22]. We are interested in modifications to spherically symmetric black holes driven by the Lorentz violating KR VEV. In a Lorentz invariant theory, the Kalb–Ramond field minimally coupled to gravity yields to an axion hairy black hole that deforms the event horizon into a naked singularity [24].

[a] e-mail: euclides.silva@ufca.edu.br (corresponding author)

Here we assume a spacelike Kalb–Ramond with constant squared norm VEV and non-minimally coupled with the Ricci tensor. As a result, we found an exact spherically symmetric and static modified black hole solution. In addition to the Schwarszchield $1/r$ solution, we obtained a power-law correction of form $\Upsilon/r^{2\lambda}$, where $\lambda$ and $\Upsilon$ are Lorentz violating (LV) parameters.

The work is organized as the following. In Sect. 2, we present a short review on the spontaneous symmetry breaking mechanism for the Lorentz symmetry driven by the Kalb–Ramond field. Moreover, we define the non-minimal coupling between the KR VEV and the Ricci tensor, and we choose the vacuum configuration. In Sect. 3 we obtain an exact solution for the modified Einstein equation found in Sect. 2. By varying the parameter $\lambda$, we analyse the modifications on the event horizon, relating them to known solutions, such as the charged black hole and the Schwarszchild–De Sitter solution. The vacuum configurations are analysed considering the behaviour of the LV source with respect to the energy conditions and the asymptotic properties of the source. In addition, the effects of this modified geometry on the black hole thermodynamics is obtained by means of the tunneling method. In Sect. 4, the modified gravitational potential is obtained, and an upper bound for the Lorentz violating parameter $\Upsilon$ is obtained using the precession period of Mercury as a test. Finally, in Sect. 5, final comments and perspectives are outlined. Throughout the text, we adopt the metric signature $(-,+,+,+)$.

## 2 The Kalb–Ramond model for spontaneous Lorentz symmetry breaking

In this section, we present the Kalb–Ramond VEV and specify the coupling of this background field with gravity.

The Kalb–Ramond field is a tensorial field arising from the bosonic spectrum of string theory [21]. It can be represented by a 2-form potential $B_2 = \frac{1}{2}B_{\mu\nu}dx^\mu \wedge dx^\nu$ whose field strength is given by $H_3 = dB_2$, or $H_{\lambda\mu\nu} \equiv \partial_{[\lambda}B_{\mu\nu]}$ in coordinates.

Inspired in the gravitational sector of the SME, we consider a self-interacting potential for the Kalb–Ramond field [22]. Assuming a potential of form $V = V(B_{\mu\nu}B^{\mu\nu} \pm b_{\mu\nu}b^{\mu\nu})$ with a non-vanishing (VEV) $< B_{\mu\nu} > = b_{\mu\nu}$ which defines a background tensor field, the Lorentz symmetry is spontaneously broken by the Kalb–Ramond self-interaction. Note that the dependence of the potential on $B_{\mu\nu}B^{\mu\nu}$ is required in order to maintain the theory invariant upon observer local Lorentz transformations. Further, the potential breaks the gauge invariance $B_2 \to B_2 + d\Lambda_1$, where $\Lambda_1$ is an arbitrary 1-form [22].

Let us consider the action for a self-interacting Kalb–Ramond field non-minimal coupled with gravity in the form [22]

$$\mathscr{S}_{KR}^{nonmin} = \int e\, d^4x \left[ \frac{R}{2\kappa} - \frac{1}{12}H_{\lambda\mu\nu}H^{\lambda\mu\nu} \right. \\ \left. - V(B_{\mu\nu}B^{\mu\nu} \pm b_{\mu\nu}b^{\mu\nu}) \right. \\ \left. + \frac{1}{2\kappa}\left(\xi_2 B^{\lambda\nu}B^\mu{}_\nu R_{\lambda\mu} + \xi_3 B^{\mu\nu}B_{\mu\nu}R\right) \right], \tag{1}$$

where $\xi_2$ and $\xi_3$ are non-minimal coupling constants (with dimensions $[\xi]=L^2$), $e$ is the metric determinant and $\kappa = 8\pi G$ is the gravitational coupling constant. Note that the non-minimal coupling enables a derivative interaction of the metric with the KR VEV.

Since we are interested in the effects of the background Kalb–Ramond VEV on the gravitational field, we consider the KR field in its vacuum configuration, i.e., $B_{\mu\nu}B^{\mu\nu} = b_{\mu\nu}b^{\mu\nu}$. In flat spacetimes, the Lorentz violating VEV $b_{\mu\nu}$ is considered constant, e.g., $\partial_\rho b_{\mu\nu} = 0$ [22]. That condition yields to a constant norm $b^2 = \eta^{\mu\nu}\eta^{\alpha\beta}b_{\mu\alpha}b_{\nu\beta}$ and allows to define Lorentz violating coefficients throughout the spacetime using $b_{\mu\nu}$ [22]. Furthermore, a constant $b_{\mu\nu}$ yields to a vanishing KR field strength $H_3 = db_2$ [22]. Thus, in flat spacetime the VEV KR is assumed to the a constant tensor with vanishing Hamiltonian. In curved spacetimes, a straightforward extension can be obtained assuming that $\nabla_\rho b_{\mu\nu} = 0$ [16]. Such condition guarantees that the KR field strength and the corresponding Hamiltonian vanish [16]. Another covariant definition of the KR VEV is furnished by assuming that $b_{\mu\nu}$ has constant norm $b^2 = b_{\mu\nu}b^{\mu\nu}$ [15]. That condition is equivalent to $b^{\mu\nu}\nabla_\rho b_{\mu\nu} = 0$ and ensures a vanishing potential [15]. In this work we assume that the KR VEV $b_{\mu\nu}$ has a constant norm and a vanishing Hamiltonian.

As performed in the Ref. [22], we use the antisymmetry of $b_{\mu\nu}$ to rewrite it as $b_{\mu\nu} = \tilde{E}_{[\mu}v_{\nu]} + \epsilon_{\mu\nu\alpha\beta}v^\alpha \tilde{B}^\beta$, where the background vectors $\tilde{E}_\mu$ and $\tilde{B}_\mu$ can be interpreted as pseudo-electric and pseudo-magnetic fields, respectively, and $v^\mu$ is a timelike 4-vector. The pseudo-fields $\tilde{E}_\mu$ and $\tilde{B}_\mu$ are spacelike, i.e., $\tilde{E}_\mu v^\mu = \tilde{B}_\mu v^\mu = 0$. Thus, the KB VEV yields two background vector instead of only one produced by the bumblebee VEV [15,16]. Moreover, the decomposition of the KR VEV in pseudo-electric and pseudo-magnetic fields is interesting since the static and modified black hole solution we find in next section has one configuration resembling a charged black hole.

In this work, we consider a pseudo-electric configuration of form

$$b_2 = -\tilde{E}(x^1)\, dx^0 \wedge dx^1. \tag{2}$$

The KR VEV *ansatz* in Eq. (2) can be rewritten as $b_2 = d\tilde{A}_1$, where $\tilde{A}_1 = \tilde{A}_0(x^1)dx^0$ can be interpreted as a pseudo-vector potential and $\tilde{E} = -\partial_1 A_0$. Since $db_2 = 0$, the VEV has a vanishing field strength $H_3$ and then a vanishing Hamiltonian. The function $\tilde{E}(x^1) = -b_{01}$ will be determined by the condition $b^2$ constant in the next section.

The constancy of $b^2$ turns the Lagrangian term $\xi_3 B^{\mu\nu} B_{\mu\nu} R$ into $\xi_3 b^2 R$, which can be absorbed into a redefinition of variables. Thus, by varying Eq. 1 with respect to the metric, the modified Einstein equations are

$$G_{\mu\nu} = \kappa T_{\mu\nu}^{\xi_2}, \tag{3}$$

with $G_{\mu\nu} = R_{\mu\nu} - \frac{1}{2} R g_{\mu\nu}$ and

$$T_{\mu\nu}^{\xi_2} = \frac{\xi_2}{\kappa} \bigg[ \frac{1}{2} g_{\mu\nu} B^{\alpha\gamma} B^\beta{}_\gamma R_{\alpha\beta} - B^\alpha{}_\mu B^\beta{}_\nu R_{\alpha\beta}$$
$$- B^{\alpha\beta} B_{\mu\beta} R_{\nu\alpha} - B^{\alpha\beta} B_{\nu\beta} R_{\mu\alpha}$$
$$+ \frac{1}{2} D_\alpha D_\mu (B_{\nu\beta} B^{\alpha\beta}) + \frac{1}{2} D_\alpha D_\nu (B_{\mu\beta} B^{\alpha\beta})$$
$$- \frac{1}{2} D^2 (B^\alpha{}_\mu B_{\alpha\nu}) - \frac{1}{2} g_{\mu\nu} D_\alpha D_\beta (B^{\alpha\gamma} B^\beta{}_\gamma) \bigg]. \tag{4}$$

Therefore, the non-minimal coupling yields to a source modifying the field equation by new derivative terms. In the next section, we seek modifications of the spherically symmetric spacetime.

## 3 Spherically symmetric solutions of Kalb–Ramond black-hole

We consider a static and spherically symmetric vacuum spacetime solution. One thus adopts the metric, as given by the line element,

$$ds^2 = -A(r)dt^2 + B(r)dr^2 + r^2 d\theta^2 + r^2 \sin^2\theta d\phi^2. \tag{5}$$

The KR VEV *ansatz* takes the form $b_2 = -\tilde{E}(r) dt \wedge dr$, where $b_{tr} = -\tilde{E}$. Since $b^2 = g^{\mu\alpha} g^{\nu\beta} b_{\mu\nu} b_{\alpha\beta}$, the Kalb–Ramond VEV *ansatz* given by Eq. (2) has a constant norm $b^2$ with the metric (5), provided that

$$\tilde{E}(r) = |b| \sqrt{\frac{A(r) B(r)}{2}} \tag{6}$$

where $b$ is a constant. Note that the function $E(r)$ in Eq. (6) defines a background radial pseudo-electric static field $\tilde{E}^\mu = (0, \tilde{E}, 0, 0)$, consistent with the spheric and static spacetime symmetry. Indeed, the background vector $\tilde{E}^\mu$ is orthogonal to both the timelike $t^\mu = (\partial/\partial t)^\mu$ and spacelike $\psi^\mu = (\partial/\partial \phi)^\mu$ Killing vectors responsible for the static and spheric symmetries [16].

Rewriting the modified Einstein Eq. (3) as

$$R_{\mu\nu} = \xi_2 \bigg[ g_{\mu\nu} b^{\alpha\gamma} b^\beta{}_\gamma R_{\alpha\beta} - b^\alpha{}_\mu b^\beta{}_\nu R_{\alpha\beta}$$
$$- b^{\alpha\beta} b_{\mu\beta} R_{\nu\alpha} - b^{\alpha\beta} b_{\nu\beta} R_{\mu\alpha} + \frac{1}{2} D_\alpha D_\mu (b_{\nu\beta} b^{\alpha\beta})$$
$$+ \frac{1}{2} D_\alpha D_\nu (b_{\mu\beta} b^{\alpha\beta}) - \frac{1}{4} D^2 (b^\alpha{}_\mu b_{\alpha\nu}) \bigg], \tag{7}$$

and using the metric *ansatz* (5), we obtain the system of equations

$$\left(1 - \frac{\lambda}{2}\right) R_{tt} = 0 \tag{8}$$

$$\left(1 - \frac{\lambda}{2}\right) R_{rr} = 0 \tag{9}$$

$$R_{\theta\theta} = \frac{\lambda r^2}{2} \left( \frac{R_{tt}}{A(r)} - \frac{R_{rr}}{B(r)} \right). \tag{10}$$

$$R_{\phi\phi} = \sin^2\theta R_{\theta\theta}, \tag{11}$$

where $\lambda := |b|^2 \xi_2$. Since the components of Ricci tensor are

$$R_{tt} = \frac{A''}{2B} - \frac{A'}{4B} \left( \frac{A'}{A} + \frac{B'}{B} \right) + \frac{A'}{rB},$$

$$R_{rr} = -\frac{A''}{2A} + \frac{A'}{4A} \left( \frac{A'}{A} + \frac{B'}{B} \right) + \frac{B'}{rB}, \tag{12}$$

$$R_{\theta\theta} = 1 - \frac{1}{B} - \frac{r}{2B} \left( \frac{A'}{A} - \frac{B'}{B} \right), \tag{13}$$

then Eqs. (8) and (9) yields to

$$A(r) = \frac{1}{B(r)}, \tag{14}$$

for $\lambda \neq 2$. Substituting the constraint Eq. (14) in Eq. (10) yields to the equation

$$\frac{r^2 \lambda}{2} A'' + (\lambda + 1) r A' + A - 1 = 0, \tag{15}$$

whose solution is

$$A(r) = 1 - \frac{R_s}{r} + \frac{\Upsilon}{r^{\frac{2}{\lambda}}}, \tag{16}$$

where $R_s = 2GM$ is the usual Schwarzschild radius and $\Upsilon$ is a constant (with dimensions $[\Upsilon]=L^{\frac{2}{\lambda}}$) that controls the Lorentz violation effects upon the Schwarzschild solution.

Therefore, the Lorentz violation trigged by the Kalb–Ramond VEV produces a power-law hairy black hole of form

$$ds^2 = -\left[ 1 - \frac{R_s}{r} + \frac{\Upsilon}{r^{\frac{2}{\lambda}}} \right] dt^2$$
$$+ \left[ 1 - \frac{R_s}{r} + \frac{\Upsilon}{r^{\frac{2}{\lambda}}} \right]^{-1} dr^2 + r^2 d\theta^2 + r^2 \sin^2\theta d\phi^2. \tag{17}$$

The modified black hole solution (17) has two Lorentz violating parameters, $\lambda$ and $\Upsilon$. The former has a definition $\lambda = |b|^2 \xi_2$ in terms of the norm of the VEV and the coupling constant $\xi_2$, whilst the latter appears as a integration constant. For a given $\Upsilon$ in the limit $\lambda \to 0$, i.e., for $|b|^2 \to 0$ or $\xi_2 \to 0$, we recover the usual Schwarzschild metric, as expected. A finite $\lambda$ reflects a relation between the VEV and the coupling constant, in the form $b^2 = \frac{\lambda}{\xi_2}$. Indeed, since the Lorentz violating effects on the gravitational field are supposed to be small, it is expected that the coupling constant to be small, as well. On the other hand, the Lorentz violating is expected to occur near the Planck scale and then, the VEV $b^2$ which triggers the Lorentz violating couplings can also be expected to be of same magnitude. Accordingly, a spontaneous breaking violation of the Lorentz symmetry enables both a large VEV and a small coupling constant as possible configurations. Unlike the mass, there is no Newtonian analogue to determine $\Upsilon$. In this work, we explore the properties of modified black holes for different $\lambda$ configurations and by studying some effects due to the LV we estimate some upper bounds for $\Upsilon$.

In addition, the Kretschmann scalar has the form

$$R_{\alpha\beta\mu\nu}R^{\alpha\beta\mu\nu} = \frac{12R_s^2}{r^6} - \frac{8r^{\frac{2}{\lambda}+1}R_s\Upsilon\left(1+\frac{3}{\lambda}+\frac{2}{\lambda^2}\right)}{r^{\frac{4}{\lambda}}r^6}$$
$$+ \frac{4\Upsilon^2\left(1+\frac{5}{\lambda^2}+\frac{4}{\lambda^3}+\frac{4}{\lambda^4}\right)}{r^{\frac{4}{\lambda}}r^4}, \quad (18)$$

and then, the LV modifications can not vanish by a coordinate change. In fact, the LBS solution Eq. (17) differs from the the Lorentz invariant KR solution [24] and from the LSB bumblebee black hole solutions in Ref. [15] and Ref. [16].

An interesting result occurs for $\lambda = |b|^2\xi_2 = -1$. In this case, $A(r) = 1 - \frac{R_s}{r} + \Upsilon r^2$, which is similar to the Schwarzschild-de Sitter (SdS) solution. Thus, the background LV Kalb–Ramond vev can be interpreted as a source for the cosmological constant $\Lambda$. The small value of the LV coefficient $\Upsilon$ provides a tiny cosmological constant, as observed.

For $\lambda = 1$, the black hole geometry (17) resembles the charged Reissner-Nordstrom solution [25,26]. However, using solution (17), the pseudo-electric field is actually constant $E(r) = \frac{|b|}{\sqrt{2}}$ and radial. This result is consistent with the asymptotic flat spacetime with a spacelike LV background field. Nevertheless, a constant electric field is inconsistent with a field created by a localized charge. Therefore, $\Upsilon$ can not be identified with a charge and it represents a LV hair of the black hole.

### 3.1 Horizons

The modified black hole solution Eq. (17) has a true singularity at $r = 0$ given by the divergence of the Kretschmann scalar (18). Nevertheless, such singularity is surrounded by a modified event horizon, thereby satisfying the cosmic censorship conjecture.

Once we have obtained the black hole solution (17), we can find the event horizons by assuming $A(r) = 0$ in (16), leading to

$$r^{\frac{2}{\lambda}} - R_s r^{\frac{2}{\lambda}-1} + \Upsilon = 0. \quad (19)$$

Note that in the local Lorentz invariant regime, i.e., for $\lambda \to 0$ or $\Upsilon \to 0$ we obtain only one horizon, $r = R_s$. Since Eq. (19) can not be solved exactly for an arbitrary $\lambda$, we assume some values for $\lambda$ and study how the KR VEV modifies the event horizons for those configurations.

For $\lambda = 1$ there are two horizons given by

$$r_\pm = \frac{R_s}{2}\left(1 \pm \sqrt{1 - \frac{4\Upsilon}{R_s^2}}\right). \quad (20)$$

Assuming $\Upsilon \ll R_s^2$, the two horizons have the form

$$r_+ \approx R_s - \frac{\Upsilon}{R_s} - \frac{\Upsilon^2}{R_s^3}, \quad (21)$$

$$r_- \approx \frac{\Upsilon}{R_s} + \frac{\Upsilon^2}{R_s^3}, \quad (22)$$

where $\lim_{R_s \to \infty} r_+ = \infty$ and $\lim_{R_s \to \infty} r_- = 0$. Therefore, the LV produces a new inner horizon $r_-$ and reduces the outer (Schwarzschild) horizon by $r_+ = R_s - r_-$. That property differs the LSB KR black hole from the bumblebee black hole found in Ref. [15] whose (Schwarzschild) horizon is kept unchanged. In Ref. [16] the authors found a LBS bumblebee black hole with only one modified horizon.

The structure of the LBS KR event horizon is rather different from the Lorentz invariant axion-KR solution [24]. In fact, the KR field turns the event horizon into a naked singularity [24], whereas the LBS KR modifies and produces new horizons.

By increasing the power which the LV term decays, the correction of the horizons decreases. For $\lambda = \frac{2}{3}$ there are three roots in Eq. (19), but only one real root. For the physical solution the changes in the usual Schwarzschild radius has the form

$$r_h \approx R_s + \mathbf{O}\left(\frac{\Upsilon^2}{R_s^6}\right). \quad (23)$$

Since no sign of Lorentz violation was found up to date, the LV parameter $\Upsilon$ is supposed to be small compared to a power of $R_s$. In the next section we establish an upper bound to $\Upsilon$ by considering the effects of the modified geometry of this LSB black hole on test particles.

### 3.2 Energy conditions

Once we studied the geometric properties of the solution (17), let us analyse the features of the KR VEV viewed as the source for the modified Einstein equation. Assuming that the stress-energy tensor (4) has an anisotropic fluid form $(T^\mu{}_\nu)^{\xi_2} = (-\rho, p_1, p_2, p_3)$, where $\rho$ is the energy density and the $p_i$ are the pressures, the source satisfies

$$\rho = \left(\frac{2}{\lambda} - 1\right)\frac{\Upsilon}{r^{\frac{2}{\lambda}+2}}, \qquad (24)$$

and the following equations of state:

$$p_1 = -\rho, \qquad (25)$$
$$p_2 = p_3 = \frac{\rho}{2}, \qquad (26)$$

Thus, a radial KR VEV yields to an anisotropic source. For $\lambda \to 0$ both the energy density and the anisotropic pressures vanish, as expected for the Lorentz invariant regime.

For $\Upsilon \geq 0$ and $\lambda$ in the interval $0 < \lambda \leq 2$, the energy density and pressure components satisfy $\rho \geq 0$, $p_1 \leq 0$ and $\rho + \sum p_i \geq 0$ and then, the weak and strong energy conditions hold [27]. As $\lambda \to 0$ the source vanishes keeping the energy conditions valid. For $\lambda = 2$ the LV source has vanishing components, regardless the value of $\Upsilon$. Therefore, the solution for $\lambda = 1$ resembling a charged black hole satisfies the energy conditions above. Yet, unlike the electromagnetic field, the KR VEV has a non-vanishing trace $\rho + \sum p_i = \rho \geq 0$.

For $\lambda \leq 0$ the weak and strong energy conditions are satisfied provided that $\Upsilon \leq 0$. Therefore, the solution for $\lambda = -1$ satisfies the energy conditions above only for a negative cosmological constant. Nonetheless, the analogy between the AdS-Schwarzschild and the $\lambda = -1$ can not be further extended since the KR VEV has an anisotropic source.

In spite of the KR VEV (6) has a vanishing Hamiltonian, the stress energy tensor $(T^\mu{}_\nu)^{\xi_2}$ (4) has non-vanishing components due to the non-minimal coupling to the Ricci tensor. Therefore, $(T^\mu{}_\nu)^{\xi_2}$ represents the Lorentz violating modifications on the Einstein equation. The components in Eqs. (24), (25) and (26) show that the geometry modifications, for $0 < \lambda \leq 2$ and $\Upsilon \geq 0$ fall off as we go far from the black hole. For $\lambda > 2$ the modifications are still localized near the black hole, though the KR VEV violates the weak energy condition.

The change on the behaviour of the LV black hole (17) from $\lambda = 1$ to $\lambda = -1$ reflects the shift on the source expressed by Eqs. (24), (25) and (26). Upon this change the energy density and pressures violates the weak energy condition and they turn out to be constant throughout spacetime. Accordingly, the LV parameters $\lambda$ and $\Upsilon$ determine the source phases and the corresponding equation of state. The gravitational phase transition is a common feature of modified gravitational theories, notably in in Lorentz violating theories [17] and in higher derivative gravities [28,29]

### 3.3 Temperature

In order to obtain the Hawking temperature for the black hole characterized by the metric (17), we will employ the Hamilton-Jacobi formalism to the tunneling approach [30–34]. In this method, the event horizon is treated as a potential barrier such that the particles created near the horizon can escape from the black hole through quantum tunneling. The method consists of computing the probability of tunneling. For this, we will consider only events near the horizon and radial trajectories, such that we can solve this in $t-r$ plane.

We consider the scalar perturbation from a massive scalar field $\phi$ around a black hole background. The equation of motion of this perturbation is the Klein–Gordon equation

$$\hbar^2 g^{\mu\nu} \nabla_\mu \nabla_\nu \phi - m^2 \phi = 0, \qquad (27)$$

where $m$ is the mass associated with the field $\phi$.

For a spherically-symmetric metric in Eq. (5) we obtain the following result after spherical harmonics decomposition

$$-\partial_t^2 \phi + \Lambda \partial_r^2 \phi + \frac{1}{2}\partial_r \Lambda \partial_r \phi - \frac{m^2}{\hbar^2} A(r)\phi = 0, \qquad (28)$$

where $\Lambda = A(r)B^{-1}(r)$.

By interpreting the field $\phi$ as a semi-classical wave function associated with the particles created in the black hole, we can solve the Eq. (28) through the WKB method which consists of using the following *ansatz* [30–34]

$$\phi(t,r) = \exp\left[-\frac{i}{\hbar}\mathcal{I}(t,r)\right]. \qquad (29)$$

Expanding (29) for the lowest order in $\hbar$, one has

$$(\partial_t \mathcal{I})^2 - \Lambda(\partial_r \mathcal{I})^2 - m^2 A(r) = 0, \qquad (30)$$

such that the Eq. (30) is the Hamilton–Jacobi equation with $\mathcal{I}$ playing the role of relativistic action. Since the metric Eq. (5) is stationary, we will look for particle-like solutions of (30) in the form [30–34]

$$\mathcal{I}(t,r) = -\omega t + W(r), \qquad (31)$$

where the $\omega$ is a constant of motion which can be interpreted as the energy of the emitted radiation. Putting the solution (31) in the Eq. (30) we obtain the following differential equation for the spatial part of the action:

$$(W')^2 = \frac{\omega^2 - m^2 A(r)}{(A(r))^2}, \qquad (32)$$

where the prime denotes derivative with respect to the radial coordinate. Also, we are using the relation $A(r) = B^{-1}(r)$

(see Eq. (14)). A straightforward integration yields

$$W(r) = \pm \int^r \frac{dr'}{A(r')} \sqrt{\omega^2 - m^2 A(r')}, \quad (33)$$

The $+$ and $-$ signs represent outgoing and ingoing particle solutions, respectively. Since we are interested in particles emitting radiation when crossing the event horizon, let us looking at the outgoing solution. Now, we take the approximation of the function $A(r)$ near the event horizon $r_+$,

$$A(r) = A(r_+) + A'(r_+)(r - r_+) + \cdots, \quad (34)$$

and the Eq. (33) takes the form

$$W = \int_0^{+\infty} \frac{dr}{A'(r_+)} \frac{\sqrt{\omega^2 - m^2 A'(r_+)(r - r_+)}}{(r - r_+)}, \quad (35)$$

in which the limits of integration were chosen so that the particle crosses the horizon $r = r_+$.

To proceed, we need to evaluate the last integral which has a simple pole at $r = r_+$. If we choose the prescription for the pole $r - r_+ \to r - r_+ - i\epsilon$, the residue theorem yields

$$W = \frac{2\pi i \omega}{A'(r_+)} + \text{(real contribution)}. \quad (36)$$

The tunneling probability of a particle escape of the black hole is given by [30–34]

$$\Gamma \sim \exp(-2\Im(\mathscr{I})) = \exp\left[-\frac{4\pi\omega}{A'(r_+)}\right], \quad (37)$$

where we note that $\Im \mathscr{I} = \Im W$.

Comparing Eq. (37) with the Boltzmann factor $e^{-\omega/T}$, we obtain the Hawking temperature of the black hole:

$$T_H = \frac{\omega}{2\Im(\mathscr{I})} = \frac{A'(r_+)}{4\pi}. \quad (38)$$

In our case, the radius of the horizon is given by (20), and the above results provide

$$T_H = \frac{R_s \left(\sqrt{R_s^2 - 4\Upsilon} + R_s\right) - 4\Upsilon}{\pi \left(\sqrt{R_s^2 - 4\Upsilon} + R_s\right)^3}, \quad (39)$$

where we assume $\lambda = 1$. As we can see, in the limit $\Upsilon \to 0$, we recover the usual Schwarzschild temperature for the classical black hole ($T_s = 1/4\pi R_s$, assuming that $R_s = 2GM$ is the Schwarzschild radius).

It is convenient to write the Hawking temperature for small values of $\Upsilon$ ($\Upsilon \ll R_s^2$). This approximation leads to a temperature

$$T_H \approx \frac{1}{4\pi R_s} - \frac{\Upsilon^2}{4\pi R_s^5}. \quad (40)$$

Note that the first term on the right in Eq. (40) is the Hawking temperature for Schwarzschild black hole. The second term represents the leading order correction due to LSB and implies that the LSB KR black hole obtained in this work is colder than the Schwarzschild black hole. This interesting feature could be considered an observational discrepancy between LSB and LI black holes and had been obtained for the bumblebee [35] and the regular [36] black holes.

## 4 Classical effects

In this section we study the effects of the Kalb–Ramond LBS black hole solution Eq. (17) on massive classical test particle. We consider only the gravitational effects upon the particle and neglect the coupling between the particle and the background KR VEV.

For a massive particle, the 4-velocity $u^\mu = \frac{dx^\mu}{d\tau}$ satisfies

$$g_{\mu\nu}(x) u^\mu u^\nu = -1. \quad (41)$$

Despite the presence of a LSB KB field, the static and radial VEV configuration in Eq. (2) preserves the time and isotropic symmetries of the black hole. Thus, the timelike Killing vector $t^\mu = \left(\frac{\partial}{\partial t}\right)$ and the spacelike Killing vector $\psi^\mu = \left(\frac{\partial}{\partial \phi}\right)$ provide two constants of motions, namely

$$E = \left(1 - \frac{R_s}{r} + \frac{\Upsilon}{r^{\frac{2}{\lambda}}}\right) \frac{dt}{d\tau}, \quad (42)$$

$$L = r^2 \frac{d\phi}{d\tau}, \quad (43)$$

which for massive particles they are the conserved energy and angular momentum per unit mass of the particle, respectively.

In terms of the energy in (42) and of the angular momentum (43), the constrain (41) can be rewritten as

$$\frac{E}{2} = \frac{1}{2}\left(\frac{dr}{d\tau}\right)^2 + V_{eff}(r), \quad (44)$$

where the effective potential $V_{eff}$ is given by

$$V_{eff}(r) = \frac{1}{2} - \frac{R_s}{2r} + \frac{L^2}{2r^2} - \frac{L^2 R_s}{2r^3} + \frac{\Upsilon}{2r^{\frac{2}{\lambda}}} + \frac{L^2 \Upsilon}{2r^{\frac{2}{\lambda}+2}}. \quad (45)$$

Accordingly, the LSB geometry induces power-law short-range forces whose strength depends on the LV parameter. Note that for $\lambda \to 0$ the LV pontential corrections vanish.

4.1 Perihelion precession

Let us now study the modifications driven by the LSB geometry on the bound orbits. In special, we consider the effects upon the advance of the perihelion and find an upper bound for the LV parameter.

By considering $\frac{dr}{d\tau} = \frac{dr}{d\phi}\frac{L}{r^2}$ and setting $x = r^{-1}$, the constrain in Eq. (44) leads to the following equation

$$\frac{d^2x}{d\phi^2} + x = \frac{R_s}{2L^2} + \frac{3R_s}{2}x^2 - \frac{\Upsilon}{\lambda L^2}x^{\frac{2}{\lambda}-1} - \left(\frac{1}{\lambda}+1\right)\Upsilon x^{\frac{2}{\lambda}+1} \quad (46)$$

Let us analyse the modifications on Eq. (46) due to the LBS. For a fixed $\Upsilon$, as $\lambda \to 0$ the corrections vanish provided that $0 < x \ll 1$, i.e., for $r \gg R_s$. Thus, the corrections on bound orbits external to the black hole are small, as expected. In addition, two interesting configurations are given by $\lambda = 1$ and $\lambda = 2/3$. The former yields to a LV correction in the linear term of Eq. (46) and a cubic term, whereas the latter generates a modification into quadratic term of Eq. (46) and a further quartic term.

Now consider the Eq. (46) for $\lambda = 1$ The ratio of the first two terms on the right-hand side of the Eq. (46) is small (of the order of $10^{-7}$ for the case of Mercury). The cubic term would make this ratio even smaller, since $\Upsilon$ is presumably small. Then, defining the parameter $\chi = \frac{3R_s^2}{4L^2}$ and neglecting the cubic term, the (46) takes the form

$$\frac{d^2x}{d\phi^2} + \left(1 + \frac{\Upsilon}{L^2}\right)x = \frac{R_s}{2L^2} + \chi\left(\frac{2L^2}{R_s}\right)x^2. \quad (47)$$

Let us look for a perturbed solution of form $x = x_0 + \chi x_1$ where $x_0$ is the Newtonian solution and $\chi x_1$ is a small deviation. Neglecting the second order terms for the parameter $\chi$, we obtain the following equation

$$\frac{d^2x_0}{d\phi^2} + \left(1 + \frac{\Upsilon}{L^2}\right)x_0 - \frac{R_s}{2L^2} + \chi\left(\frac{d^2x_1}{d\phi^2} + \left(1 + \frac{\Upsilon}{L^2}\right)x_1 - \frac{2L^2}{R_s}x_0^2\right) = 0. \quad (48)$$

The zeroth-order solution $x_0$ is given by

$$x_0 = \frac{R_s}{2L^2(1+\frac{\Upsilon}{L^2})}\left(1 + e\cos(\sqrt{1+\frac{\Upsilon}{L^2}}\phi)\right), \quad (49)$$

where $e$ is the eccentricity of an ellipse and the first-order solution is

$$x_1 = \frac{R_s}{2L^2\left(1+\frac{\Upsilon}{L^2}\right)^2}\left[\frac{1+e^2}{2\left(1+\frac{\Upsilon}{L^2}\right)} + \frac{e}{\sqrt{1+\frac{\Upsilon}{L^2}}}\phi\sin\left(\sqrt{1+\frac{\Upsilon}{L^2}}\phi\right) - \frac{e^2}{6(1+\frac{\Upsilon}{L^2})}\cos\left(2\sqrt{1+\frac{\Upsilon}{L^2}}\phi\right)\right]. \quad (50)$$

which for these three terms obtained only $\frac{e}{\sqrt{1+\frac{\Upsilon}{L^2}}}\phi\sin(\sqrt{1+\frac{\Upsilon}{L^2}}\phi)$ is important in the correction of $x_0$, since it accumulates over successive orbits, i.e., after each revolution it gets larger and larger, unlike the other two terms.

Thus, the general solution has the form

$$x \approx \frac{R_s}{2L^2(1+\frac{\Upsilon}{L^2})}\left[1 + e\cos[(1-\beta)\sqrt{1+\frac{\Upsilon}{L^2}}\phi]\right], \quad (51)$$

where

$$\beta = \frac{3R_s^2}{4L^2(1+\frac{\Upsilon}{L^2})^{\frac{3}{2}}}. \quad (52)$$

Therefore, the orbit period with the LV correction that will be given by

$$\Phi = 2\pi + \Delta\Phi_{GR} + \Delta\Phi_{LV}, \quad (53)$$

where $\Delta\Phi_{GR} = \frac{3\pi R_s}{(1-e^2)a}$ is GR correction with $a$ being the semi-major axis of the orbital ellipse.

The contribution of the Lorentz violation of the model is given by

$$\Delta\Phi_{LV}^{\lambda=1} = -\frac{2\pi\Upsilon}{R_s(1-e^2)a} \quad (54)$$

Similarly, we can perform the same analysis for $\lambda = 2/3$ which contributes with a quadratic correction due to LV. Likewise the $\lambda = 1$ configuration, we disregard the quartic term due to its small contribution and we consider only the correction on the quadratic term. Accordingly, we find that the correction in the period is given by

$$\Delta\Phi_{LV}^{\lambda=2/3} = -\frac{6\pi\Upsilon}{R_s(1-e^2)^2 a^2}, \quad (55)$$

which is smaller than the $\lambda = 1$ correction. Indeed, that effects is expected since the greater $\frac{2}{\lambda}$ is the less is the LV correction upon the Schwarzschild solution, for a fixed $\Upsilon$. Therefore, we expect that the higher is $\frac{2}{\lambda}$, the lower its contribution to the gravitational corrections.

The correction Eq. (54) allow us to estimate an upper bound for the Lorentz violation parameter $\Upsilon$ from the perihelion shifts for the orbit of the planet Mercury. In fact, as predicted by general relativity, the perihelion advance $\Delta\Phi_{GR} = 42.9814''/C$ (in arcseconds per century) and the observational error is $e = 0.003''/C$, according to the most up to date observations found in Refs. [37,38]. Thus, we obtained $\Upsilon_{\lambda=1} < 2.8 \times 10^{-3} km^2$. For a small astrophysical black hole with ten times the solar mass, the shift in the event horizon is of order

$$\frac{\Upsilon^{\lambda=1}}{(R_s)^2} \approx 2.0 \times 10^{-6}. \quad (56)$$

It is worthwhile to mention that for $\lambda = 1$ the LV parameter $\Upsilon$ has mass dimension $M^2$. The relativaly large value of $\Upsilon$ for $\lambda = 1$ suggests that LBS modifications are suppressed by smaller values of $\lambda$.

## 5 Final remarks and perspectives

In this work, we obtained a spherically symmetric and static solution of gravity non-minimally coupled to a VEV of the Kalb–Ramond field. The self-interaction potential breaks the KR gauge invariance and produces a VEV background tensor field which violates the local Lorentz symmetry.

The vacuum background tensor was chosen in order to vanish the self-interaction potential and the Kalb–Ramond hamiltonian. Further, the VEV is perpendicular to the timelike and spacelike Killing vectors and then, the Lorentz violation preserves the static and spheric symmetries of the gravitational vacuum. By assuming a non-minimal coupling between the KR VEV and the Ricci tensor, we found a power-law correction to the Schwarzschild solution. The black hole solution has two parameters controlling the Lorentz violation, $\lambda$ and $\Upsilon$. For $\lambda = 1$, the LSB solution exhibits two horizons, likewise a charged black hole. Since no charge or angular momentum is present, the solution represents a Lorentz violating hairy black hole. The analysis of the equation of the state and the energy conditions of the Lorentz violating source reveals that the weak energy condition is valid for $0 \leq \lambda \leq 2$ and $\Upsilon > 0$.

The radius of the outer horizon is the Schwarzschild radius minus the inner horizon. For $\lambda = -1$ the LSB solution resembles a Schwarzschild-Anti-De Sitter black hole, with the Lorentz symmetry parameter $\Upsilon$ being proportional to the cosmological constant. However, such analogy is not complete due the anisotropic source. Further, for $\Upsilon < 0$ the source violates the weak energy condition. The higher is the $\lambda$ the lower is the correction to the horizon radius and then, we focus on the $\lambda = 1$ configuration for the sake of simplicity. It is worthwhile to mention that, unlike the LSB black holes generated by the bumblebee field, the KR LSB solution found not only modifies the usual Schwarzschild event horizon but also produces additional horizons.

The Lorentz violation also modifies the black hole temperature. We employed the tunneling method to derive the Berkenstein–Hawking (BH) temperature. It turns out that the Lorentz violating parameter $\Upsilon$ reduces the BH temperature by a term proportional to $\Upsilon^2$. Similar results were also obtained in other modified black holes, such as the LSB bumblebee [35] and the regular black hole [36].

At the classical level, the LSB solution found yields to an additional gravitational potential whose power depends on the Lorentz violating parameter $\lambda$. For $\lambda = 1$, we obtained an upper bound for the LV parameter $\Upsilon$ by studying the correction of the precession period of the planet Mercury. From this bound, we can estimate that for a peculiar black hole with ten times the solar mass, the shift in the outer event horizon is about 20 mm compared to 30 km of the Schwarzschild radius. The scale of the upper bound we found suggests that the LV gravitational effects are suppressed by higher powers, i.e., for smaller $\lambda$. Thus, the precesssion period analysis restricts $\lambda$ to $0 < \lambda \ll 1$. A numerical analysis to obtain the temperature and the period correction for an arbitrary $\lambda$ is an important perspective to achieve a better bounds for $\lambda$ and $\Upsilon$.

As possible extensions of the present work, we point out the stability analysis of the solution found by considering the corrections due to fluctuations of the KR field around the VEV. The analysis for magnetic-monopole KR VEV configurations, such as $b_2 = \tilde{B}(r)r^2 \sin\theta d\theta \wedge d\phi$, could exhibits a lower power-law correction. Further, other non-trivial KR VEV with non-vanishing hamiltonian might provide richer solutions. Moreover, wormhole and regular solutions can also be found. The effects of a coupling between the KR field and the Riemann tensor is another important development. The phase transition analysis by means of the black hole thermodynamics is another noteworthy prospect.

**Acknowledgements** The authors thank the Conselho Nacional de Desenvolvimento Científico e Tecnológico (CNPq), Grants n$^\circ$ 312356/ 2017-0 (JEGS), n$^\circ$ 305678/2015-9 (RVM) and n$^\circ$ 308638/2015-8 (CASA) for financial support.

**Data Availability Statement** This manuscript has associated data in a data repository. [Authors' comment: The article "Modified Black hole solutions with a background Kalb Ramond field", https://doi.org/10.1140/epjc/s10052-020-7902-1 has no data associated in a data repository. Only a draft version is public available on arXiv:1911.10296 [gr-qc].]